# Remote Life Detection Criteria, Habitable Zone Boundaries, and the Frequency of Earth-like Planets around M and Late-K Stars

James F. Kasting[*], R. Kopparapu, R. M. Ramirez, and C. E. Harman

Department of Geosciences, Penn State University, University Park, PA 16802, USA

*Corresponding author: Ph 814-865-3207, email jfk4@psu.edu

**Abstract**

**The habitable zone (HZ) around a star is typically defined as the region where a rocky planet can maintain liquid water on its surface. That definition is appropriate, because this allows for the possibility that carbon-based, photosynthetic life exists on the planet in sufficient abundance to modify the planet's atmosphere in a way that might be remotely detected. Exactly what conditions are needed, however, to maintain liquid water remains a topic for debate. Historically, modelers have restricted themselves to water-rich planets with $CO_2$ and $H_2O$ as the only important greenhouse gases. More recently, some researchers have suggested broadening the definition to include arid, "Dune" planets on the inner edge and planets with captured $H_2$ atmospheres on the outer edge, thereby greatly increasing the HZ width. Such planets could exist, but we demonstrate that an inner edge limit of 0.59 AU or less is physically unrealistic. We further argue that conservative HZ definitions should be used for designing future space-based telescopes, but that optimistic definitions may be useful in interpreting the data from such missions. In terms of effective solar flux, $S_{eff}$, the recently recalculated HZ boundaries are: recent Venus—1.78, runaway greenhouse—1.04, moist greenhouse—1.01, maximum greenhouse—0.35, early Mars—0.32. Based on a combination of different HZ definitions, the frequency of potentially Earth-like planets around late-K and M stars observed by Kepler is in the range of 0.4-0.5.**

Abbreviations: HZ, habitable zone; $S_{eff}$—stellar flux divided by the solar flux at Earth's orbit; $T_{eq}$, planetary equilibrium temperature; $T_*$, stellar effective temperature; RV, radial velocity; $\eta_{Earth}$, the frequency of Earth-like planets around other stars; TPF—Terrestrial Planet Finder

\body

As the appearance of this special issue of PNAS confirms, the search for exoplanets is by now well underway; indeed, it is one of the hottest research areas in all of astronomy. At the time of this writing, 428 exoplanets have been identified by ground-based radial velocity (RV) measurements and 279 planets have been discovered by the transit technique [1] and confirmed by various other methods, including RV. In addition, more than 3000 "planet candidates" have been identified by the Kepler mission [2]. Most of these planets are either too large or too close to their parent star to have any chance of harboring life. But both astronomers and the general public are ultimately interested in identifying potentially habitable planets and in searching their atmospheres spectroscopically for evidence of life. This leads immediately to the question of how life can be recognized remotely, along with the related question of what are the conditions needed to support it. Liquid water is often mentioned as a prerequisite for life, and we will argue below that this restriction is appropriate for the astronomical search for life. Some researchers have questioned this assumption, though [3], and so we begin by explaining why the presence of liquid water is so important in the search for life on planets around other stars.

**Criteria for remote life detection: why liquid water matters**

Biologists sometimes debate what constitutes life, but one definition that most of them accept is the following [4,5]: "Life is a self-sustained chemical system capable of undergoing

Darwinian evolution". That definition appeals to biologists because it is general and because it can be tested in the laboratory. Astronomers, however, are interested in looking for life on planets around other stars by performing remote sensing of the planets' atmospheres, so to them the biologists' definition of life is not particularly useful. Instead, what they need is a way to recognize life from a great distance. It was realized many years ago that the best way to do this is by looking for the byproducts of metabolism. As early as 1965, Lederberg [6] suggested that the best remote signature of life was evidence for extreme thermodynamic disequilibrium in a planet's atmosphere (but see criticism of that idea below). In that same year, Lovelock [7] put forward a more specific remote signature, namely, the simultaneous presence of $O_2$ and a reduced gas such as $CH_4$ or $N_2O$ in a planet's atmosphere. On Earth, all three of these gases are produced mostly by organisms, and they are many orders of magnitude out of equilibrium with each other, so the fact that they coexist is indeed strong evidence that our planet is inhabited.

This life detection criterion, however, is much less general than is often supposed, as demonstrated by this counter-example. Suppose that a planet's atmosphere was found to have high concentrations of both CO and $H_2$. These gases are quite far out of thermodynamic equilibrium at room temperature, as free energy considerations strongly favor the reaction: $CO + 3 H_2 \rightarrow CH_4 + H_2O$. Thus, a naive interpretation might be that the observed disequilibrium was evidence for life. But CO-rich, $CH_4$-poor, atmospheres can be produced by impacts [8] or by photolysis of $CO_2$ in cold, dry, low-$O_2$ atmospheres [9]. The presence of life would likely destabilize such a CO-rich atmosphere, as either methanogens would consume it [10] or, alternatively, acetogens would use it to produce acetate [11]. So, the criterion of extreme thermodynamic equilibrium as a biomarker is directly contradicted.

It is also easy to demonstrate that the converse of this statement is not true: the absence of extreme thermodynamic disequilibrium should *not* be taken as evidence that life does not exist on the planet. To illustrate why this is so, consider the Earth's atmosphere and biosphere prior to the origin of oxygenic photosynthesis. Models suggest that organisms in such an anaerobic biosphere would combine $CO_2$ and $H_2$ to make $CH_4$, either directly (by methanogenesis) or indirectly (by anoxygenic photosynthesis, followed by fermentation and methanogenesis) [11]. In either case, the early biosphere would have been driving the atmosphere *towards* thermodynamic equilibrium, not away from it, because the reaction

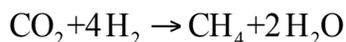
$$CO_2 + 4H_2 \rightarrow CH_4 + 2H_2O$$

is energetically downhill at room temperature.

Recently, Seager et al. [12,13] have proposed a systematic classification of potential biomarker gases, which is useful. They divide them into three categories: Type 1 biomarkers are redox gradient by-products, such as $CH_4$ in the case just described. They are produced when organisms take advantage of existing thermodynamic free energy gradients to power their metabolisms. These biomarkers are inherently equivocal, because abiotic processes might also synthesize them, provided that suitable kinetic pathways exist. $CH_4$, for example, can be formed by serpentinization of ultramafic rocks by water containing dissolved $CO_2$ [14]. Similarly, Seager et al. suggest that $NH_3$ might be a Type I biomarker on a cold "Haber world" with a dense $N_2$-$H_2$ atmosphere, because it is difficult to produce $NH_3$ by gas-phase photochemistry. But $NH_3$ can also be produced by photolytically catalyzed of reaction of $N_2$ with $H_2O$ on surfaces containing $TiO_2$ [15], and $NH_3$ might be synthesized from dissolved $N_2$ and $H_2$ within submarine hydrothermal systems on a planet with a highly reduced mantle, like Mars [16]. Type II biomarkers are biomass building by-products, such as the $O_2$ produced from oxygenic photosynthesis. But this could conceivably be the only useful Type II biomarker, as anoxygenic photosynthesis based on $H_2S$ or $Fe^{++}$ yields the non-gaseous by-products, elemental sulfur and $Fe_2O_3$, respectively, and other forms of anoxygenic photosynthesis yield highly soluble by-products like sulfate and nitrate, or $H_2O$ itself [12]. Type III biomarkers are secondary metabolic products, such as DMS (dimethyl sulfide), OCS, $CS_2$, $CH_3Cl$, and higher hydrocarbons such as isoprene. These gases, however, are produced in sufficiently small amounts and/or are photolyzed sufficiently rapidly that they are not expected to build up to observable concentrations in a planet's atmosphere. So, the number of potentially detectable biomarker molecules may be relatively limited.

Although the Seager et al. scheme is a nice way of classifying biomarker gases, it is not clear that it really changes the fundamental nature of remote life detection. One still needs to observe something in a planet's atmosphere that cannot be explained by abiotic processes. Significantly, nearly all of the hypothetical planets that these authors discuss possess liquid water. They do discuss Titan, but they argue that it requires unphysically high biomass amounts based on terrestrial analogies. We would go farther than this. Searching for life on our own Solar System's Titan is a fascinating endeavor, because perhaps we will be surprised and find that life exists there. But remote sensing of Titan-like planets around other stars is not likely to convince many biologists that life exists there, *unless* life has already been discovered on Titan itself.

One final thought about liquid water planets should be noted. Many authors have suggested that life could conceivably be present on a planet with a subsurface liquid water reservoir. These conjectures can eventually be tested by in-situ measurements on bodies within our Solar system, such as Mars or Jupiter's moon, Europa. But for exoplanets, in-situ sampling remains indefinitely out of reach, and the ability of a subsurface biosphere to modify a planet's atmosphere in a detectable way remains to be demonstrated. Thus, in looking for signs of life on exoplanets, it makes sense to search for planets like Earth that have abundant liquid water at their surfaces. In other words, the habitable zone (HZ), as conventionally defined, is precisely where we should focus our attention.

**Habitable zone boundary definitions**

*Inner edge*. Two different types of climate catastrophe can be triggered if an Earth-like planet's surface becomes too warm. If the surface temperature exceeds the critical temperature for water (647 K for pure $H_2O$), the entire ocean should evaporate. This is termed a *runaway greenhouse*. Alternatively, Goldblatt et al. [17] define a runaway greenhouse as an atmosphere in which the absorbed solar flux exceeds the outgoing thermal-IR flux. But these two definitions are equivalent, in practice, because an atmosphere that is out of flux balance in this way 'runs away' to supercritical surface temperatures. A runaway greenhouse could occur at a surface temperature less than 647 K on a planet that had less water than Earth; the critical temperature applies on Earth because the pressure of a fully vaporized ocean (270 bar) exceeds the critical pressure for water (220.6 bar). Runaway greenhouse atmospheres have finite lifetimes, because once the upper atmosphere becomes water-rich, the water can be lost in a few tens of millions of years by photodissociation followed by escape of hydrogen to space [18,19].

However, a planet can also lose its water to space at much lower temperatures. Once the surface temperature exceeds ~340 K for a 1-bar Earth-like atmosphere, the $H_2O$ saturation mixing ratio at the surface exceeds 0.2. This leads to a large increase in tropopause height and an accompanying large increase in stratospheric $H_2O$ [18,19]. The presence of gaseous $H_2O$ above the altitude at which it can condense then overcomes the diffusion limit on hydrogen escape [20], so the water can be rapidly photodissociated and lost. This is termed the *moist greenhouse* limit. Finally, a third, empirical limit can be defined based on the observation that Venus has no liquid water, even though its initial water endowment may have been high, based on the planet's high D/H ratio [21]. Indeed, Venus had apparently already lost any surface water it might have possessed by 1 billion years (b.y) ago (the approximate age of its most recent resurfacing [22]), at which time the Sun was ~8% less bright than today [23]. This third estimate is termed the *recent Venus* limit. It implies that the inner edge of the HZ is at least 4% farther out than Venus' orbital distance of 0.72 AU.

According to Kasting et al. [23], the three estimates for the effective solar flux, $S_{eff}$, at the HZ inner edge for a G2 star like the Sun were: moist greenhouse, or "water loss"—1.10, runaway greenhouse—1.41, and recent Venus—1.76. In terms of orbital distance, these correspond to values of 0.95 AU, 0.84 AU, and 0.75 AU, respectively, for the present solar luminosity. Note that the HZ boundaries are not defined in terms of the planet's equilibrium temperature, $T_{eq}$, even though that concept is sometimes erroneously applied in the astrophysics literature. To calculate an equilibrium temperature, one must have knowledge of a planet's albedo. The value of 0.31 for present Earth [24], though, is not applicable to a planet near the inner or outer edge of the HZ. In the model of [23], planets near the moist greenhouse limit were predicted to have albedos of ~0.2, while planets near the maximum greenhouse limit had predicted albedos of ~0.4. So, instead, the preferred way to define the HZ boundaries is in terms of the incident stellar flux at the top of the planet's atmosphere. These boundaries change for stars of different spectral types, because the calculated albedos of moist greenhouse and maximum greenhouse planets become lower (higher) as the star's radiation is shifted towards the red (blue). The change in albedo is caused partly by Rayleigh scattering, which increases strongly at shorter wavelengths, and partly to the increase in near-IR absorption by $H_2O$ and $CO_2$ as the star's spectral peak shifts to these wavelengths.

*Outer edge.* Following [23], Earth-like planets are assumed to be volcanically active and to have substantial supplies of carbon in the form of $CO_2$ and carbonate rocks. $CO_2$ is removed from the atmosphere by silicate weathering, followed by deposition of carbonate sediments on the seafloor. This process depends on the amount of rainfall, which slows as the climate cools [25]; hence, atmospheric $CO_2$ should build up on planets that receive less stellar insolation than does Earth. At some point, however, $CO_2$ begins to condense out of the atmosphere, as it does today on Mars. This led, historically, to two theoretical estimates for the HZ outer edge. The *first condensation* limit is where the process of $CO_2$ condensation first begins. That limit, however, is no longer considered valid, because $CO_2$ ice clouds are now thought to generally warm a planet's surface [26,27], and so we ignore it from this point on. The still-accepted estimate for the location of the outer edge is termed the *maximum greenhouse* limit (but see arguments below for how this distance might be extended). As its name implies, this is the distance at which the warming by $CO_2$ maximizes, so the solar flux required to sustain a 273 K surface temperature is at a minimum. $CO_2$ partial pressures above the value at this limit lead to surface cooling, because the increased albedo caused by Rayleigh scattering from the $CO_2$ outstrips the increase in the $CO_2$ greenhouse effect. Finally, an empirical outer edge for the HZ can be defined based on the observation that Mars appears to have had liquid water flowing on its surface at or before 3.8 b.y. ago, when the solar flux was some 25 percent lower [23]. Some authors have argued that this does not imply a long-lived warm climate, because the fluvial features can be produced in other ways, and we return to this question below. For now, we simply assume that early Mars was within the HZ. According to [23], the three estimates for $S_{eff}$ at the HZ outer edge around a G star were: first $CO_2$ condensation—0.53, maximum greenhouse—0.36, early Mars—0.32. The corresponding orbital distances were 1.37 AU, 1.67 AU, and 1.77 AU, respectively, for present solar luminosity.

Other authors have proposed modifications to these HZ boundaries. Selsis et al. [28] included clouds in the model of [23] and calculated boundaries for 50% and 100% cloud cover. This procedure is somewhat questionable, as the model of [23] already included clouds implicitly as part of the surface albedo. Not surprisingly, for 100% cloud cover, the inner edge for a Sun-like star moves in to ~0.5 AU, or $S_{eff}$ = 4. But this result is unrealistic, because condensation clouds are not expected to have such high fractional cloud cover and because it is in direct conflict with the existence of a desiccated Venus at 0.72 AU ($S_{eff}$ = 1.9). The 50% cloud cover model produces more believable results, as it predicts an inner edge at 0.76 AU—very close to the recent Venus limit. But the cloud layer in this model was placed at an arbitrary height, with an arbitrary optical depth, and the influence of the cloud on the planet's greenhouse effect was neglected. So, it is not obvious that this estimate represents any real improvement over the empirical recent Venus limit.

More recently, other authors have proposed modifications to the HZ boundaries based on suggestions of different types of habitable planets and/or the presence of additional greenhouse gases (see [29] for a recent review). Specifically, Zsom et al. [30] have used 1-D climate model calculations to argue that the inner edge of the HZ could be as far in as 0.5 AU around a Sun-like star for planets with low relative humidities (RH). For example, their nominal, low-RH planet has a constant tropospheric RH of 0.01. By comparison, Earth's RH varies from ~0.8 at the surface to ~0.1 in the upper troposphere [31]. Such a low-RH planet could, in principle, avoid the strong water vapor feedback that makes the runaway greenhouse happen. But this model fails to consider energy balance at the planet's surface. The net absorbed (solar + thermal-IR) flux at the surface must equal the convective (latent + sensible) heat flux [32]. Because the latent heat flux is a strong function of RH, it can be readily shown that liquid water would be unstable on the surface of a warm, low-RH planet if the surface is all at the same temperature. A detailed explanation is given in the SI Appendix. Zsom et al. argue that there would be cold parts of the planet where water would condense, but then their 1-D model is not really appropriate. Consequently, their predictions about the location of the HZ inner edge are suspect.

Climate on low-RH humidity planets has been studied previously using 3-D models. Abe et al. [33] used such a model to show that hot, rocky planets with small water endowments and low obliquities could conceivably remain habitable in their polar regions. Such planets would resemble the planet *Dune* in Frank Herbert's famous science fiction novel by that title; hence, the name "Dune-like" planet has stuck. Abe et al. placed the inner edge of the HZ at $S_{eff} = 1.7$, well inside the runaway or moist greenhouse limit for a water-rich planet, but still outside the recent Venus limit. Such planets would be consistent with surface energy balance, but they might be unrealistic for another reason. Some of the moisture originating from polar oases on Dune-like planets would presumably fall as rain on drier portions of the planet's surface. Once there, it might well become incorporated into minerals, e.g., clays, as water of hydration. More water could conceivably be tied up in seafloor basalts, as happens on Earth today during circulation of seawater through midocean ridges. Earth's seafloor is eventually subducted, so this water is carried down towards the mantle. This process would have removed about half an ocean of water in 2.5 billion years, were water not being continually returned to the surface by outgassing [34]. A Dune-like planet would presumably have a much smaller water inventory, and so the removal time might be considerably shorter. In any case, even if Dune-like planets do exist, the inner edge of the HZ is still well outside our empirical recent Venus limit.

The outer edge of the HZ could conceivably be further out than the estimates mentioned above as a consequence of warming by $CO_2$ ice clouds [26] or by additional greenhouse gases. $CO_2$ ice clouds apparently have only a modest warming effect on climate, based on new 3-D studies of early Mars [35]. But the addition of significant concentrations of molecular hydrogen, $H_2$, can extend the outer edge of the HZ dramatically [36,37]. We ourselves have suggested that early Mars could have been warmed by this mechanism [38]. $H_2$ is perfect for this task, for two reasons: 1) Its (collision-induced) absorption extends over the entire thermal-infrared spectrum [39], and 2) it condenses only at extremely low temperatures. Realizing this, Pierrehumbert and Gaidos [37] showed that a 3-Earth-mass planet with a 40-bar captured $H_2$ atmosphere could conceivably remain habitable out to 10 AU around a Sun-like star. Stevenson [36] had previously demonstrated that such planets could remain habitable even if they were not attached to stars at all, provided that they had sufficient internal heat. While such planets might exist, we should think carefully about whether this should influence the design of a future habitable-planet-seeking telescope. The contrast ratio between the Earth and the Sun is ~$10^{10}$ in the visible [40], which is already considered difficult to achieve. A planet of the same size and albedo located at 10 AU would have a contrast ratio 100 times larger. Would we really want to spend the extra money required to develop a starlight suppression system with this efficiency? Our dollars might be better spent ensuring that the telescope can find planets more similar to Earth.

This raises a more general question: Should estimates of HZ width be conservative or optimistic? In our view, both types of estimates are useful, but for different purposes. In designing a space telescope like NASA's proposed Terrestrial Planet Finder (TPF) (also known as New Worlds Observer, NWO), one of the key questions is how big should it be. A bigger telescope should be more capable, but it is also likely to cost more and may take longer to be approved and built. TPF telescope size is dictated primarily by two factors: 1) One needs to be able to observe enough target stars to have a good chance of finding Earth-like planets. The TPF-C STDT report [40] suggested that the planet expectation value should be 3, implying that the chances of finding at least one Earth candidate is >95%. 2) One needs to be able to distinguish the light from the planet from that of the surrounding exozodi dust. Exozodi levels are not yet well constrained, but using our own zodi background as a proxy suggests that the minimum telescope diameter is 4 m [41].

For constraint (1) above, a key parameter is the value of $\eta_{Earth}$—the fraction of Sun-like stars that have at least one rocky planet within their HZ. If $\eta_{Earth}$ is high, then fewer stars need to be searched. The target stars should therefore be closer, and so the angular separation, $\theta$, between them and their potentially habitable planets should be larger. The theoretical diffraction limit is given by: $\theta \gtrsim \lambda/D$, where $\lambda$ is the wavelength at which one is observing and $D$ is the telescope diameter. Real coronagraphs operate only outside of some multiple of this angle; for example, TPF-C was designed with an inner working angle of 4 $\lambda/D$ [40]. Closer target stars thus translate into a smaller telescope diameter. The TPF-C committee assumed that $\eta_{Earth} = 0.1$; thus, they designed the telescope to be able to search 30 full habitable zones (or, more accurately, 60 half habitable zones around 60 nearby target stars). To do this, they required an 8×3.5 m elliptical mirror. Recent estimates of $\eta_{Earth}$ for late-K and M stars suggest that $\eta_{Earth}$ could be significantly higher than this (see discussion below). If Sun-like stars follow this same pattern, then perhaps a 4-m diameter TPF telescope will be sufficient.

Estimating $\eta_{Earth}$ requires that one employ a specific definition of the habitable zone. In doing this, we suggest that one should use a conservative HZ definition, for example, the moist greenhouse limit on the inner edge and the maximum greenhouse limit on the outer edge. That way, the $\eta_{Earth}$ estimate will also be conservative, and the telescope will be at least as big as it needs to be to find the requisite number of planets. Most importantly, it would be a mistake to take the outer edge at 10 AU for a Sun-like star, because then $\eta_{Earth}$ might be calculated as unity, and the telescope might be correspondingly undersized. Using a more optimistic inner edge for the HZ is less of a problem, because this would require a smaller inner working angle, which itself would place constraints on telescope diameter. But it might turn out that a larger and more expensive telescope would be needed to find Dune-like planets close to their stars, and one would again have to ask whether this additional expenditure would be worthwhile.

Once we have constructed such a telescope and are using it to make observations, then adopting a more liberal HZ definition makes sense. At that point, it would be prudent to examine all rocky planets within the most optimistic HZ, so that one would not accidentally overlook possible Dune planets or $H_2$-rich super-Earths in systems where they could be observed. Distant, $H_2$-rich planets would likely be hard to see, as pointed out above, and their potential biomarkers might be difficult to interpret; however, there is no reason that one could not look for them with TPF, even if it was designed to search for more conventional, Earth-like planets.

**Revised HZ boundaries and calculations of $\eta_{Earth}$**

Recently, our group has re-derived HZ boundaries using a revised and improved 1-D climate model. Our new model [42] includes updated absorption coefficients for both $CO_2$ and $H_2O$, with the latter being based on line parameters from the relatively new HITEMP database.

This database includes significantly more weak $H_2O$ lines than does the older HITRAN database, which is used in most current climate models. This change is particularly important in the visible/near-UV spectral region, where our new model predicts much stronger absorption of stellar radiation, thereby lowering planetary albedos and moving the inner edge of the HZ outward.

Figure 1 shows the new HZ boundaries around stars of different spectral types in terms of effective stellar flux, $S_{eff}$, which is the way that we recommend that they be implemented in practice. Parametric formulae for these limits are given by eq. (2) of [42], with coefficients listed in their Table 3. The formulae are expressed in terms of the parameter $T_* = T_{eff} - 5780\,\text{K}$, where $T_{eff}$ is the star's effective radiating temperature. For a star like the Sun, with $T_{eff}$ = 5780 K, the new HZ boundaries are given in units of $S_{eff}$ as: recent Venus—1.78, runaway greenhouse—1.04, moist greenhouse—1.01, maximum greenhouse—0.35, early Mars—0.32. The recent Venus and early Mars limits are the same as in [23], because the planets are still at the same orbital radii and because solar evolution models have not changed significantly since that time. We note that the new, theoretical inner edge limits are perilously close to Earth's present orbit; however, these limits are almost certainly overly pessimistic and can be improved on by using 3-D climate models, as discussed below. Figure 2 shows the new HZ boundaries in terms of distance. For simplicity, only the moist greenhouse and maximum greenhouse limits are shown. This diagram was calculated for stars at the beginning of their main sequence lifetimes, and it is only valid at that time, because all main sequence stars brighten as they age. Fig. 2 shows that on a log distance scale the HZ is slightly broader for late-type stars than for Sun-like stars. The reason is that the planet's albedo depends on the spectrum of its parent star, and this dependence is stronger at the HZ outer edge than at the inner edge because Rayleigh scattering is so important in dense $CO_2$ atmospheres. So, if planets around late-type stars are distributed geometrically in orbital distance, as they are in our own Solar System, one might expect $\eta_{Earth}$ to be ~25% larger for M stars than for G stars.

Geometrical spacing of planets is predicted by some theories of planetary formation. . Dynamical analyses [43-45] have identified a well-defined stability boundary in multi-planet systems. This result led to the "packed planetary systems" (PPS) hypothesis [46-48], which proposes that planetary systems form in such a way that they are filled to capacity: adding even one additional planet would create an unstable system. Although this is a plausible hypothesis, it will have to be verified by future observations.

Differences in HZ boundary definitions account for a large part of the differences in recent estimates for $\eta_{Earth}$ from both Kepler and RV data. Dressing and Charbonneau [49] calculated $\eta_{Earth} = 0.15^{+0.13}_{-0.06}$ for M stars, based on the first 6 quarters of Kepler data. Their analysis used the moist greenhouse and $CO_2$ 1$^{st}$ condensation limits from [23]. We pointed out earlier that the latter limit is no longer considered valid. Furthermore, they only included planets up to 1.4 Earth radii ($R_{Earth}$). But planets up to ~2 $R_{Earth}$ should have masses <10 $M_{Earth}$ and thus could be rocky, although some planets could have significantly more water ice than does Earth [50]. Kopparapu [51] repeated Dressing and Charbonneau's analysis using their same dataset but applying the new moist greenhouse and maximum greenhouse limits from [42]. For planets up to 1.4 $R_{Earth}$ in size, he derived $\eta_{Earth} = 0.48^{+0.12}_{-0.24}$, rising to $0.51^{+0.1}_{-0.2}$ when he included planets up to 2 $R_{Earth}$. He then repeated the analysis again using the more optimistic recent Venus and early Mars limits and calculated $\eta_{Earth} = 0.61^{+0.07}_{-0.15}$. More recently, Gaidos [52], using the 50% cloud cover HZ boundaries of [28] and considering planets up to 2 $R_{Earth}$, derived an independent estimate of $\eta_{Earth} = 0.46^{+0.18}_{-0.15}$ based on the first 8 quarters of Kepler data. These boundaries are roughly comparable to the recent Venus and early Mars limits mentioned above; hence, his estimate for $\eta_{Earth}$ ought to be higher than Kopparapu's conservative value, but it is not. Gaidos' dataset is 30 percent longer and includes both M and late K stars, so perhaps this represents a trend towards lower $\eta_{Earth}$ around more massive stars. Finally, Bonfils et al. [53] derived $\eta_{Earth} = 0.41^{+0.54}_{-0.13}$ based on a 6-yr radial velocity study of 102 nearby southern M dwarfs, using the optimistic recent Venus/early Mars HZ limits. Thus, estimates for $\eta_{Earth}$ for stars significantly smaller than the Sun seem to be converging at values of the order of 0.4-0.5, although the error bars remain large and the different estimates are not strictly comparable because of different assumptions regarding planet size and HZ boundaries. Because the HZ for G stars is 25 percent narrower than for M stars, $\eta_{Earth}$ for G stars would be of the order of 0.3-0.35 if other factors, *e.g.* efficiency of planetary formation, planetary spacing, are equal.

**Future work**

Theoretical studies of HZ boundaries and observational determinations of $\eta_{Earth}$ are still both in their infancy. The most important theoretical question concerns the location of the HZ inner edge. The red area in Fig. 1 shows the large amount of orbital real estate between the conservative moist greenhouse limit and the optimistic recent Venus limit. The actual inner edge probably lies in between these two estimates, but where? This problem can potentially be resolved by more complicated, 3-D climate modeling. 1-D models are incapable of providing reliable answers to this question, for two reasons: 1) the relative humidity distribution is uncertain, and 2) cloud feedback is impossible to gauge effectively. The moist greenhouse calculation, as performed by [23] and [42], assumes a fully saturated troposphere. This is overly pessimistic, as Earth's present troposphere is significantly undersaturated, on average, in its upper regions. The reasons have to do with cumulus convection, especially in the tropics, where the air remains saturated within convective plumes, but then the relative humidity drops as the air spreads out at the top of the plume and descends more slowly over wider areas. The dynamics of this process can be captured in 3-D models but not in 1-D. Similarly, clouds form mostly on updrafts, where the air is cooling, and they are largely absent in regions where the air is subsiding, such as over the Sahara desert. Again, 3-D models can capture such dynamics, although even with them it is difficult because most clouds are sub-grid-scale in size. Cloud feedback has been predicted to be negative for warm, $H_2O$-rich atmospheres [19,28] because their contribution to the planetary albedo is likely to be higher than their contribution to the greenhouse effect. Strong negative cloud feedback is exhibited in a recent 3-D climate calculation of synchronously rotating M-star planets [54]. If this result is correct, this would move the HZ inner edge for M stars to $S_{eff} \cong 2$. That said, cloud feedback is actually calculated to be *positive* in many existing 3-D climate models of present Earth, because high cirrus clouds, which tend to warm the climate, are predicted to increase faster with increasing surface temperature than lower stratus clouds, which tend to cool it [55]. Pushing suitably equipped 3-D climate models up into the warm, moist greenhouse regime could help elucidate how clouds and relative humidity change with surface temperature and, thus, where the inner edge of the HZ actually lies.

In summary, much work remains to be done both to calculate reliable HZ boundaries and then to find and characterize rocky planets within them. But this work is well worthwhile, as it could eventually lead to the detection of habitable exoplanets, and perhaps to evidence of extraterrestrial life.

Acknowledgements: We thank two anonymous reviewers for constructive comments which helped us frame our arguments more precisely. Funding for this work was provided by the NASA Astrobiology Institute's Virtual Planetary Laboratory and the NASA Exobiology and Evolutionary Biology Research Program. The authors are affiliated with Penn State's Center for Exoplanets and Habitable Worlds.

**Figure Captions**

Fig. 1  Diagram showing different HZ boundaries for stars ranging in spectral type from F0 to M7. The "1st $CO_2$ condensation" limit is no longer considered valid for the outer edge, for reasons discussed in the text. Various planets within our Solar System are shown, along with selected exoplanets.

Fig. 2  Diagram showing the HZ boundaries in terms of log distance for zero-age-main-sequence stars. The moist greenhouse limit is used for the inner edge and the maximum greenhouse limit for the outer edge. Stellar luminosities and effective temperatures were taken from [56,57].